\documentclass[twoside,twocolumn]{article}

\usepackage{blindtext} % Package to generate dummy text throughout this template 

\usepackage{graphicx}
\usepackage[sc]{mathpazo} % Use the Palatino font
\usepackage[T1]{fontenc} % Use 8-bit encoding that has 256 glyphs
\usepackage{microtype} % Slightly tweak font spacing for aesthetics
\usepackage{flushend}
\usepackage[english]{babel} % Language hyphenation and typographical rules
\usepackage[hmarginratio=1:1,top=32mm,columnsep=20pt]{geometry} % Document margins
\usepackage[hang, small,labelfont=bf,up,textfont=it,up]{caption} % Custom captions under/above floats in tables or figures
\usepackage{booktabs} % Horizontal rules in tables
\usepackage{float}
\usepackage{placeins}
\usepackage[super,numbers]{natbib}
\usepackage{balance}
\usepackage[]{geometry}% http://ctan.org/pkg/geometry

\usepackage{enumitem} % Customized lists
\setlist[itemize]{noitemsep} % Make itemize lists more compact

\usepackage{glossaries}
\setacronymstyle{long-short}
\newacronym{pes}{PES}{Potential Energy Surface}
\newacronym{mep}{MEP}{Minimal Energy Path}
\newacronym{neb}{NEB}{Nudged Elastic Band}
\newacronym{cineb}{CINEB}{Climbing Image Nudged Elastic Band}
\newacronym{gsm}{GSM}{Growing String Method}
\newacronym{idpp}{IDPP}{Image Dependent Pair Potential}
\newacronym{dft}{DFT}{Density Functional Theory}
\newacronym[plural=Neural Networks]{nn}{NN}{Neural Network}
\newacronym[plural=Message Passing Neural Networks]{mpnn}{MPNN}{Message Passing Neural Network}
\newacronym[plural=Graph Neural Networks]{gnn}{GNN}{Graph Neural Network}
\newacronym{mse}{MSE}{Mean Squared Error}
\newacronym{ci}{CI}{Climbing Image}
\newacronym{ml}{ML}{Machine Learning}
\newacronym[plural=Gaussian Processes]{gp}{GP}{Gaussian Process}
\newacronym{painn}{PaiNN}{Polarizable Atom interaction Neural Network}
\newacronym{ase}{ASE}{Atomic Simulation Environment}
\newacronym{qm}{QM}{Quantum Mechanics}
\newacronym{dftb}{DFTB}{Density Functional based Tight Binding}
\newacronym{mae}{MAE}{Mean Average Error}
\newacronym{rmse}{RMSE}{Root Mean Squared Error}
\newacronym{t1x}{T1x}{Transition1x}
\newacronym{ani1x}{ANI1x}{ANI1x}
\newacronym{qm9}{QM9}{QM9}

\title{NeuralNEB -- Neural Networks can find Reaction Paths Fast \\\vspace{-15pt} \hrulefill \vspace{-15pt}} % Article title
\author{%
Mathias Schreiner\textsuperscript{\rm 1} \and 
Arghya Bhowmik\textsuperscript{\rm 2} \and 
Tejs Vegge\textsuperscript{\rm 2} \and
Peter Bjørn Jørgensen\textsuperscript{\rm 2} \and 
Ole Winther\textsuperscript{\rm 1,3,4} \and
\\  % Your name
\textsuperscript{\rm 1} \normalsize DTU Compute, Technical University of Denmark (DTU)\\
\textsuperscript{\rm 2} \normalsize DTU Energy, Technical University of Denmark (DTU)\\
\textsuperscript{\rm 3} \normalsize Department of Biology, University of Copenhagen (UCph)\\
\textsuperscript{\rm 4} \normalsize Genomic Medicine, Copenhagen University Hospital, Rigshospitalet\\
\vspace{-35pt}
}

\usepackage{balance}
\usepackage{fancyhdr} % Headers and footers
\usepackage{titling} % Customizing the title section
\usepackage{hyperref} % For hyperlinks in the PDF
\date{} % Leave empty to omit a date
\begin{document}
\thispagestyle{empty}
\newgeometry{top=20pt,bottom=0pt,right=75pt,left=75pt}
\onecolumn
\maketitle

%\noindent \textbf{\LARGE NeuralNEB -- Neural Networks can find Reaction Paths Fast} \\
%\noindent \large{Mathias Schreiner\textsuperscript{1}, Arghya Bhowmik\textsuperscript{1}, Tejs Vegge\textsuperscript{1}, Peter Bjørn Jørgensen\textsuperscript{1}, \\Ole Winther\textsuperscript{1,2}}\\ 
%Technical University of Denmark (DTU), 2800 Lyngby, Denmark \\
%\hrulefill

\begin{abstract}
Quantum mechanical methods like \gls{dft} are used with great success alongside efficient search algorithms for studying kinetics of reactive systems. However, \gls{dft} is prohibitively expensive for large scale exploration. \gls{ml} models have turned out to be excellent emulators of small molecule \gls{dft} calculations and could possibly replace \gls{dft} in such tasks. For kinetics, success relies primarily on the models' capability to accurately predict the \gls{pes} around transition-states and \glspl{mep}. Previously this has not been possible due to scarcity of relevant data in the literature. In this paper we train state of the art equivariant \gls{gnn}-based models on around 10.000 elementary reactions from the recently published Transition1x dataset. We apply the models as potentials for the \gls{neb} algorithm and achieve a \gls{mae} of 0.13$\pm$0.03  eV on barrier energies on unseen reactions. We compare the results against equivalent models trained on QM9x and ANI1x. We also compare with and outperform \gls{dftb} on both accuracy and required computational resources. The implication is that \gls{ml} models are now at a level where they, given relevant data, aptly can be applied for studying reaction kinetics. 
\end{abstract}

\begin{figure}[h]
  \makebox[\textwidth]{%
    \begin{minipage}[t]{\textwidth}
      \centering
      \includegraphics[scale=0.85]{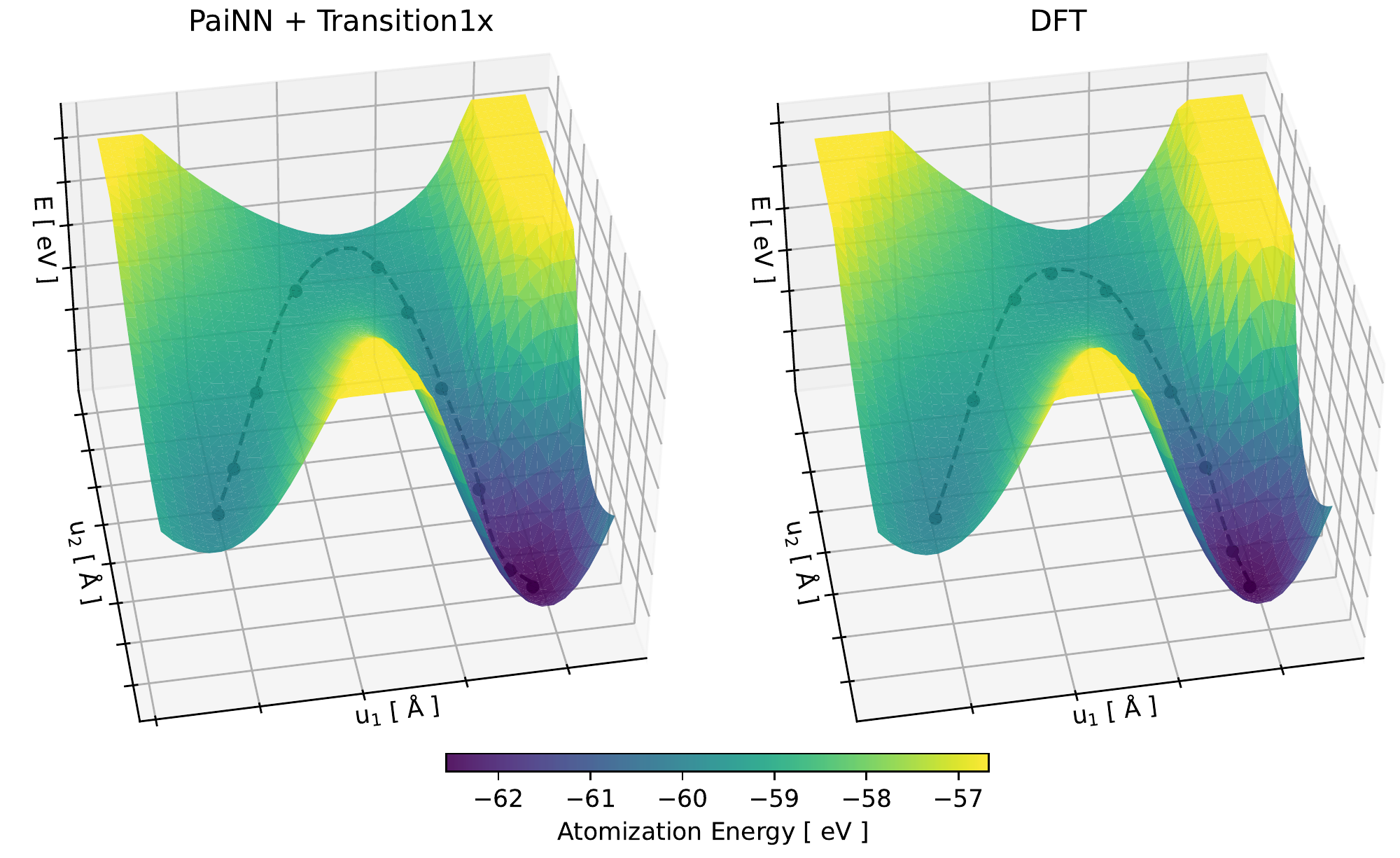}
      \caption{
      \acrfullpl{mep} found with \acrfull{neb} applying the \acrfull{gnn} architecture PaiNN trained on the Transition1x dataset and \acrfull{dft} as potentials. The \glspl{mep} are projected onto the planes in structural space, intersecting product, reactant and transition-state of the converged \glspl{mep}. The \gls{pes} has been calculated on the planes in the vicinity of the \glspl{mep} with the respective potential and is shown with colors and on the z-axis. The x and y-axes are basis vectors describing the plane. The reaction involves a H-transfer coupled with a C-C bond formation on C6H8. The reaction can be seen as a GIF by following  \href{https://figshare.com/articles/media/Reaction/20377497}{this link}.
     }
    \label{fig:frontpage}
    \end{minipage}%
  }%
\end{figure}

\newpage
\restoregeometry
\twocolumn

\section{Introduction}
% background 
\gls{ml} models and especially \glspl{gnn} \cite{gnnBacciu2019,gnnZhou2021} have turned out to be potent emulators of \gls{dft} potentials for small molecules \cite{Faber2017NNDFT, Westermayr2021ml4qm1, Campbell2020ml4qm2, behler2007generalized, Westermayr2021qm4ml4}, thanks to their remarkable ability to find complex relations in high dimensional data. They have a complexity-scaling orders of magnitudes lower than classic \gls{qm} methods, but have in recent years achieved comparable accuracy \cite{nnvsdft1,nnvsdft2,nnvsdft3,nnvsdft4,nnvsdft5}. The capability of these models is manifested by their success in tasks beyond simple prediction of molecular features such as structural optimization or studying finite-temperature dynamical properties through molecular dynamics \cite{Kaappa2021structuraloptimization, Wang2020finitetemp}. Despite their achievements, there has only been limited success in applying \gls{ml}-models as potentials for transition search algorithms. The earliest work studied simple diatomic molecule dissociation and achieved acceptable accuracy with tens of thousands of data points \cite{malshe2007theoretical}. Other works have had success by limiting their scope to studying single or few reactions but sacrificing the generality of the approach \cite{lu2018rate, young2021transferable, manzhos2020neural}. % Recent studies have predicted reaction barriers from reactant structures only \cite{heinen2021toward} and used response operator based quantum machine learning to perform transition-state-search \cite{von2022geometry}. 
Attempts to study reactive systems with \glspl{gp} \cite{Koistinen2017gp} have been successful too, but the \gls{gp} is trained on the particular atomic system, sacrificing speed for generality by requiring expensive \gls{dft} calculations at inference time. Transition-states are notoriously hard to find as there is no well-defined gradient on the \gls{pes} to guide traditional optimization algorithms towards them. A wealth of algorithms have been proposed to solve this problem -- one is the \gls{neb} \cite{neb} algorithm, which works by interpolating an initial path between reactant and product and iteratively updating it to minimize energy by using information about the \gls{pes}. It shares a common bottleneck with other transition search algorithms -- the necessity to repeatedly evaluate energy and atomic forces of molecular configurations, which is extremely costly, especially if ab-initio or electron \gls{dft} calculations are used\cite{heinen2020machine}. 

Recent advances in \gls{ml} have not alleviated the bottleneck as even modern \gls{nn} architectures have not proved proficient potential approximators for this type of application. The fault lies primarily with available data in the literature rather than the models' expressiveness \cite{anatol2020chemspace}. Most quantum mechanical datasets are focused on molecular configurations in or near equilibrium \cite{ani1Smith2017,ani1xSmith2020,gdb11.1Fink2007,gdb11.2Fink2005}. Without configurations on and around reaction pathways in the training data, \gls{ml} models cannot learn the interatomic interactions that occur during chemical reactions and cannot reliably be applied for transition-state search. 

We compare \gls{ml} models against \gls{dftb}, \cite{dftb} a fast approximation to \gls{dft} that is often used for fast screening of large quantities of configurations with an acceptable trade-off between accuracy and speed, and our models outperform \gls{dftb} with a factor five in accuracy and a factor 2.5 in CPU time.

% summary
In this work, we bridge generalization, speed, and accuracy for transition-state search by applying \gls{painn} models as surrogate potentials for \gls{dft}. We build on and showcase the utility of our previous paper \cite{Schreiner2022} where we released Transition1x, a dataset constituted by \gls{dft} calculations for 10 million molecular configurations, all sampled around reaction pathways from 10.000 elementary reactions. It is clear from the results of this paper, that for precise modeling of transition-state regions, and, consequently, transition states and barrier energies, hitherto popular benchmark datasets have had insufficient relevant data. On the other hand, training \gls{ml} potentials on the Transition1x dataset allows for accurate modeling of \glspl{pes} in transition-state regions, underlining that relevant and available data in the literature is as important as the efficiency of available models.

% training \gls{painn} \cite{painn} models on the recently published and novel Transition1x \cite{Schreiner2022} -- a dataset constituted by \gls{dft} calculations for 10 million molecular configurations, all sampled around reaction pathways from 10.000 elementary reactions -- and apply the models as potentials for running \gls{neb}. It is clear from the results that for precise modeling of transition-state regions, and, consequently, transition states and barrier energies, hitherto popular benchmark datasets have had insufficient relevant data. 

% future value
Reliable and fast analysis of reaction kinetics through \gls{ml} will bring the whole field of computational chemistry a considerable step closer to the ultimate goal, a virtual laboratory, hyper-accelerating the discovery of reaction mechanisms for synthesizing drugs and materials.

\section{Methods}
\subsection{Nudged Elastic Band}
\gls{neb} \cite{neb} is a method for finding \gls{mep} and transition-state given product and reactant of a chemical reaction. It does so by iteratively nudging an interpolated path between the reaction endpoints in the direction of the force perpendicular to the path. Once the perpendicular force converges to zero, \gls{neb} reports the maximal-energy configuration along the path as the transition-state. The path is represented by an array of molecular configurations called images, and there is no guarantee that, at convergence, the maximal energy image corresponds to the maximal energy along the path. The maximum might lie between two images. \gls{cineb} \cite{cineb} addresses this problem by letting the transition-state candidate (the maximal energy image) further maximize its energy by following the gradient on the \gls{pes} parallel to the current path between iterations. If the current path has not converged properly, the climbing image can pull the predicted \gls{mep} off the true \gls{mep} and therefore, the path is first relaxed with regular \gls{neb} before turning on \gls{cineb}. The \gls{mep} is considered converged once the maximal perpendicular force on the path is below a threshold of 0.05 eVÅ$^{-1}$. The spring constant between images on the path is set to 0.1 eVÅ$^{-2}$, and ten images are used to represent the path. 

\subsection{Initial Path Generation}
The endpoints of the reaction have to be minimized in their respective minima before running \gls{neb} -- otherwise the energetic difference between reactant and transition-state cannot be evaluated properly. A configuration is considered relaxed if the norm of the forces acting on it is below 0.01 eVÅ$^{-1}$. Once the endpoints have been minimized, the initial guess for the \gls{mep} is found by running \gls{neb} with the \gls{idpp} \cite{idpp} on a linearly interpolated path between reactant and product. \gls{idpp} is an inexpensive potential specifically designed to generate physically realistic \gls{mep} guesses for \gls{neb} at an extremely low computational cost. 

\subsection{Optimizers}
Reactants and products are relaxed using the BFGS \cite{bfgs} optimizer with $\alpha=70$ and a maximal step size of 0.03 Å in configurational space. The \gls{mep} is found with an optimizer \cite{neboptimizerMakri2019} designed to reduce the computational cost of transition-state search algorithms by applying an adaptive time step selection algorithm with $\alpha=0.01$ and rtol $=0.1$, and a preconditioning scheme to the \gls{pes} given an estimate of its curvature.  

\section{Data}
We train all models on ANI1x \cite{ani1xSmith2020}, QM9x \cite{qm9}, Transition1x \cite{Schreiner2022}. All datasets are calculated with the 6-31G(d) \cite{6-31G} basis set and $\omega$B97x \cite{wb97x} functional which has an accuracy comparable to the gold standard but expensive high-level CCSD(T) \cite{wb97xvsccsdt1Riley2010} \cite{wb97xvsccsdt2Thanthiriwatte2011} calculations. Given the compatibility of the datasets, it is possible to train on either dataset alone or combinations of them to leverage all of their strengths.

\subsection{ANI1x} 
ANI1x \cite{ani1x} aims to provide varied data of off-equilibrium molecular configurations by perturbing equilibrium configurations with pseudo molecular dynamics. The data is collected through an active learning technique called Query by Committee; an automated data diversification process that trains an ensemble (committee) of models on a dataset and accepts or rejects new proposed data based on the disagreement of models in the committee. The assumption is that if the committee disagrees the data is sufficiently different from what has already been learned, and the proposed data should be included in the analysis. The procedure for proposing data and evaluating it with the committee is cheap compared to the calculation of data using \gls{dft}. The dataset is consecutively expanded by alternating between training committees and adding new data points based on the committee uncertainty. In total, ANI1x contains force and energy calculations for approximately 5 million configurations.

\subsection{Transition1x}
We have recently published Transition1x \cite{Schreiner2022}, a dataset providing a collection of molecular configurations on and along reaction paths for approximately 10.000 reactions. The reactions consist of up to 7 heavy atoms, including C, N, and O. Transition events are rare, and it is not possible to collect sufficient data in relevant regions by simple molecular dynamics if the intention is to train \glspl{nn} models to understand chemical reactions. Transition1x addresses this problem by sampling molecular configurations around reaction pathways proposed by \gls{neb}, using \gls{dft} as potential. The procedure resulted in approximately 10 million \gls{dft} calculations that were collected and saved during the process and constitute the dataset. Transition1x is available through the repository \href{https://gitlab.com/matschreiner/Transition1x}{https://gitlab.com/matschreiner/Transition1x} which includes data loaders and scripts for downloading the dataset and generating ASE-database files.

\subsection{QM9 and QM9x}
QM9 \cite{qm9} is a dataset of 135k small organic molecules with various chemical properties that has served as the benchmark for many existing \gls{ml} methods for quantum chemistry. All molecules in QM9 are in equilibrium. We have recalculated QM9 with the 6-31G(d) basis set and $\omega$B97x functional to make it compatible with Transition1x and ANI1x, and we refer to the recalculated dataset as QM9x. Molecular configurations recalculated in the new potential are not necessarily in equilibrium as the potential shifts when changing functional and basis sets. QM9x is available through the repository \href{https://gitlab.com/matschreiner/QM9x}{https://gitlab.com/matschreiner/QM9x} which includes data loaders and scripts for downloading the dataset and generating ASE-database files.

\subsection{Models and Training}
\glspl{mpnn} \cite{mpnnGilmer2017} are a class of \glspl{gnn} \cite{gnnBacciu2019,gnnZhou2021} that build their internal graph representation by running a series of message passing steps. A single message passing step consists of two distinct operations: i) \textit{Message Dispatching}, each node computes a message given its state (and possibly information about the edge connecting to -- and the state of the receiving node) and sends it to its neighbors. ii) \textit{State Update}, incoming messages are collected with an aggregation function, and are used to simultaneously update the internal representation of all nodes. After the message-passing phase, a readout function extracts the inner representation of the nodes and computes a final feature vector of the graph for downstream tasks. In the case of molecules, interesting properties are energy and forces where conservative force fields can be computed via the back-propagation algorithm as the negative gradient of the energy wrt. coordinates of the atoms.

The \gls{painn} model \cite{painn} was used for all experiments -- it is a \gls{gnn} architecture that implements rotationally equivariant representations for prediction of tensorial properties of graph structures. We refer to the literature for further details \cite{painn}. A cut-off radius of 5 Å was used to generate the initial molecular graph. All models have three message passing steps and 256 units in each hidden layer, and are trained using the ADAM \cite{adam}  optimizer with learning rate $10^{-3}$ on training examples from QM9x, ANI1x, and Transition1x. A batchsize of 75 was used for all datasets and a maximum of $10^{6}$ training steps was allowed - however, models training on ANI1x and Transition1x reached maximal scores on validation data after around $6\cdot 10^{5}$ steps. Transition1x was stratified by reactions without attention to reaction-type. 500 reactions were set aside for testing such that no data from around any test reaction was seen during training. 90\% of the remaining data were used for training and the last 10\% were used for validation and early stopping. ANI1x was stratified by chemical formula such that test, validation and training sets consist of chemical formulas unique to that set. QM9x was split randomly. In the case of QM9x and ANI1x, 80\% of the data was used for training, 10\% for testing, and 10\% was used for validation and early stopping. In QM9x all configurations are unique as they are in distinct equilibria and can therefore be split randomly. No attention was paid to the molecular scaffold. For ANI1x, it is necessary to split on chemical formula to ensure that configurations across splits are significantly different. Each chemical formula contains similar configurations, since data is generated by randomly perturbing identical initial configurations. In Transition1x the reactions follow unique paths in chemical space, and thus configurations between reactions are different. The Transition1x dataset forms a reaction network where the product from one reaction is the reactant for the next. Therefore, it is nontrivial to split it to get no shared reactants/products across train validation and test sets. We believe that our splitting is still useful because it achieves unique reaction paths and transition states between datasets. 

\section{Results}

\begin{table*}[ht]
\begin{tabular}{|l|ll|rrr|} \hline
& \multicolumn{2}{c|}{Barrier {[} eV {]}} & \multicolumn{3}{c|}{NEB Convergence} \\ 
& \multicolumn{1}{c}{MAE} & \multicolumn{1}{c|}{RMSE} & \multicolumn{1}{c}{Rate} & \multicolumn{1}{c}{ Avg. CPU Time} & \multicolumn{1}{c|}{Avg. Iterations} \\ \hline
        ANI1x  & 0.43(1) & 0.56(3)  &71.9\% & 36s & 130.91  \\
        T1x  & \textbf{0.13(3)} & \textbf{0.24(1)}  & 80.5\% & 28s & 111.97    \\
        QM9x  & 3.4(1) & 3.6(1)  &23.4\% &\textbf{ 26s} & 110.34  \\
        DFTB  & 0.66 & 0.79  &72.0\% & 81s & 115.65 \\
        DFT  & - & - &  \textbf{84.1\%} & 12h14m43s & \textbf{100.74} \\ \hline 
\end{tabular}
\caption{
Performance of various potentials used for \acrfull{neb} when compared to \acrfull{dft}. ANI1x, Transition1x and QM9x indicate PaiNN models trained on the respective dataset. The Barrier column displays the \acrfull{mae} and \acrfull{rmse} of barrier predictions, where the individual error is the difference between the barrier as predicted when using \gls{dft} as potential vs. using the surrogate potential. The convergence rate is the percentage of reactions that converged. Average CPU time is CPU time spent per reaction. Average iterations is the average number of \acrfull{mep} updates before convergence. 
}
\label{tab:results}
\end{table*}

\begin{figure*}[ht]
    \centering
    \includegraphics[scale=0.8]{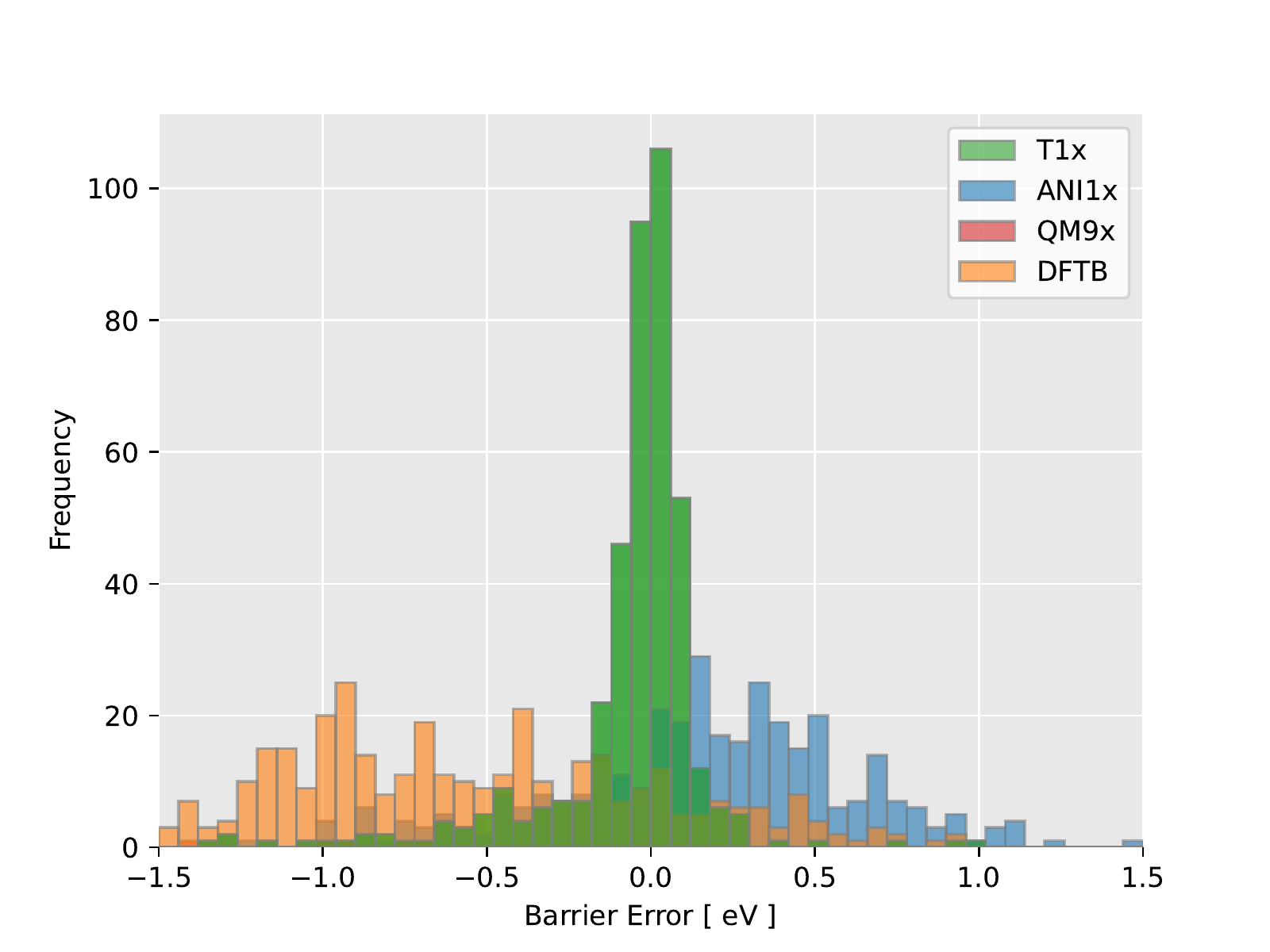}
    \caption{
    Histogram of barrier errors. The x-axis shows errors between reaction barriers calculated using \acrfull{dft} and surrogate potentials for \acrfull{neb}. The x-axis has been truncated at +/- 1.5 eV error (see appendix for the full plot). The y-axis shows the frequency of each bin. Green, red and blue display results from PaiNN models trained on Transition1x, QM9x and ANI1x, respectively. Yellow displays results from \acrfull{dftb}. The QM9x model has such a low convergence frequency, and general barrier error, that the model does not show in the plot.
    }
\label{fig:histogram}
\end{figure*}

\begin{figure*}[ht]
    \centering
    \includegraphics[scale=0.8]{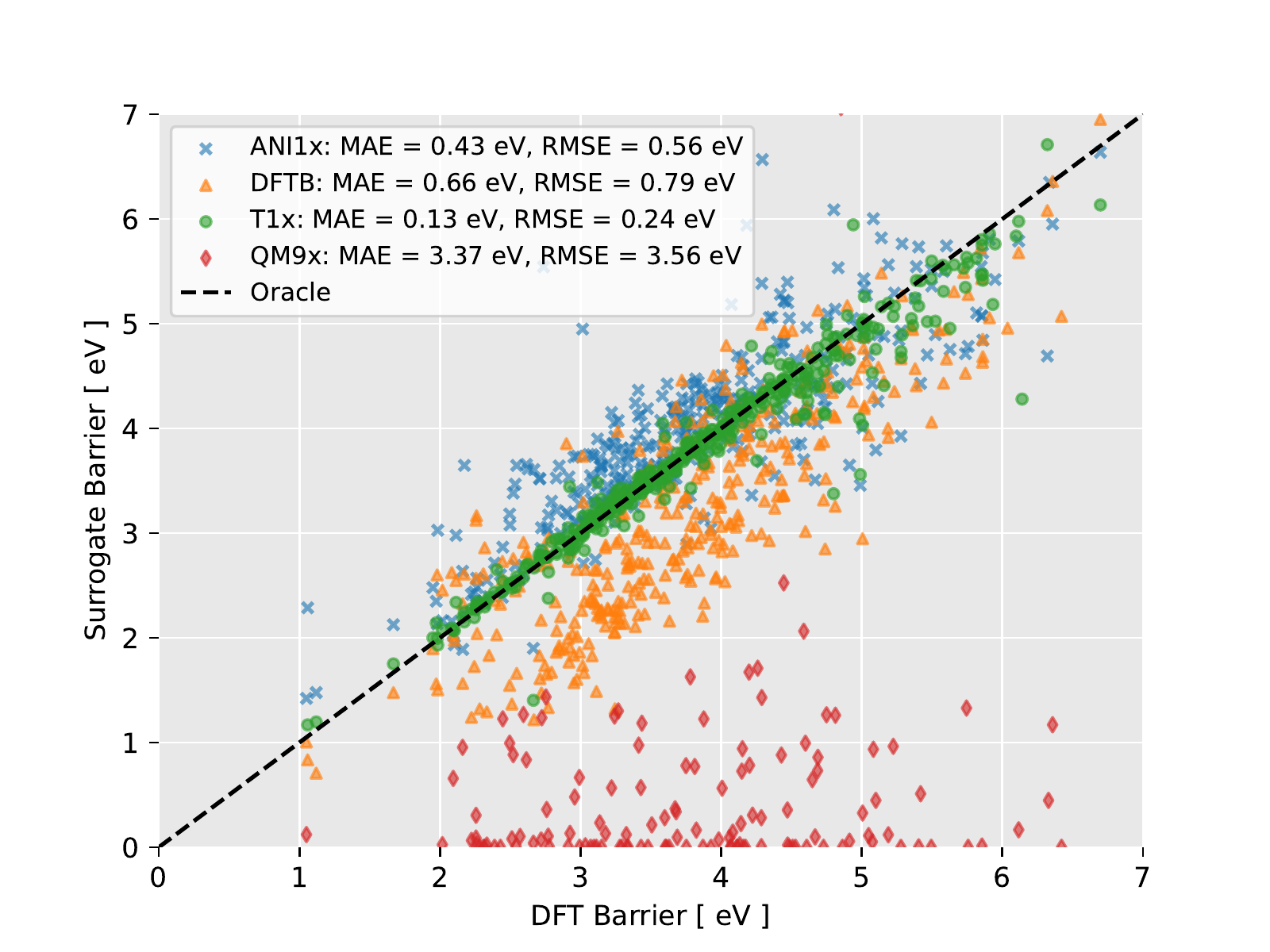}
    \caption{
    Comparison of reaction barriers found with \acrfull{neb} using \acrfull{dft} as potential on the x-axis vs. various surrogate potentials on the y-axis. Green, red and blue markers are PaiNN models trained on The Transition1x, QM9x, and ANI1x datasets respectively. Yellow is \acrfull{dftb}. Points on the dashed line have been calculated perfectly.}
\label{fig:scatter}
\end{figure*}

\begin{table*}[ht]
\begin{tabular}{|l|ccc|}
      \hline

      & MAE [ eV ] & RMSE [ eV ] & Systematic Error [ eV ]  \\ 

      \hline
ANI1x & 0.40 & 0.56 & 0.16 \\
DFTB  & 0.47 & 0.57 & -0.55 \\
Transition1x & 0.17 & 0.30 & -0.07 \\
      \hline

\end{tabular}\caption{
MAE and  \gls{rmse} of barrier errors found by PaiNN trained on Transition1x and ANI1x and \gls{dftb}, after correcting for systematic error.}
\label{tab:corrected}
\end{table*}

Table \ref{tab:results} shows the overall findings of the paper. Each row displays the performance of a surrogate potential, where datasets in the leftmost column indicate \gls{painn} models trained on the given dataset. The barrier error is the difference in barrier heights found when applying \gls{dft} as potential for \gls{neb} versus when applying the respective potential. As different initializations of parameters in equivalent architectures result in variations in the trained models, five models were trained on each dataset without changing training, testing and validation splits. These models were used as potentials for \gls{neb}, the table displays the mean and standard deviation of \glspl{mae} and \glspl{rmse} of each model. 

The best models are trained on Transition1x, with the lowest \gls{mae} and \gls{rmse} and the highest convergence ratio. The QM9x models have only seen data very close to equilibrium and have not learned the structure of the \gls{pes} between equilibria which makes it unable to converge in most cases. In general \gls{dft} performs the best in terms of convergence rate and average iterations run, but it comes at a steep price, running almost a factor 1500 times slower than the \gls{ml} potentials. \gls{dftb} is the go-to fast potential, but the models trained on Transition1x are roughly 2.5 times faster and five times as accurate.

Figure \ref{fig:frontpage}, on the frontpage, displays \glspl{mep} calculated with \gls{neb} using \gls{dft} and \gls{painn} trained on Transition1x side by side. Each \gls{mep} is projected onto a plane in configurational space spanned by the reaction's transition-state, product, and reactant. The x and y axes are basis vectors describing the plane in units of Å, and the z-axis and color-coding show the atomization energy of configurations in the plane in eV. Not only does \gls{painn} trained on the Transition1x accurately calculate the barrier energy for the reaction, but it also correctly identifies the plane spanned by the configurations defining the reaction, and calculates an almost identical \gls{pes} in the vicinity of the \gls{mep}. Each \gls{mep} is projected from a high dimensional space onto the plane, and therefore, only the atomization energy of equilibria and transition-states are shown correctly in the plot. At these points, the \gls{mep} intersects with the plane. The intermediate points have energies slightly shifted up the sides of the energy valley. The \gls{mep} does not necessarily lie in the plane, and since the \gls{mep} represents the energy valley, projecting it onto the plane, will increase the energy. The $\times$ symbols on the surfaces are projections of images predicted by \gls{neb} and the dashed lines connecting them are cubic spline interpolations. The importance of accurate predictions in the vicinity of the \gls{mep} is clear, as these calculations will guide the search for the transition-state. The Transition1x model predicts smooth and well-behaved \glspl{pes} resembling \gls{dft}.

Figure \ref{fig:histogram} and \ref{fig:scatter} tell similar stories. Figure \ref{fig:histogram} is a histogram of barrier errors where the error is the difference between activation energy found using the surrogate potential and \gls{dft}. The Transition1x model is precise and accurate, with a sharp peak around zero, whereas \gls{dftb} and ANI1x have wider spreads with means below and above zero, respectively. The QM9x model is plotted on the histogram, but due to high errors and low convergence, only a few calculated barriers fall within an error of $\pm$1.5 eV, as shown in the figure. See appendix for an equivalent figure without truncated x-axis. 

Figure \ref{fig:scatter} compares activation energies found with \gls{dft} on the x-axis with those found using various surrogate potentials on the y-axis. Each marker represents a single reaction. Predictions from the model trained on Transition1x follow the $x=y$ line with a \gls{mae} of only 0.13 eV. The QM9x model does not have a proper representation of the transition-state regions as it has not seen that type of data during training. Often, the QM9x model does not recognize nearby initial equilibria as minima on the \gls{pes}, and even before optimizing the \gls{mep}, the reaction endpoints have dropped further on the \gls{pes} to qualitatively different endpoints which results in the model calculating the \gls{mep} for a completely different reaction. The algorithm is not set up to detect this, and as long as the reaction converges, it is included in the analysis. Even when the QM9x model relaxes the endpoints of the reaction correctly, it either finds low energy shortcuts in the faulty potential or does not converge, and as a result the converged reactions are often only the energy difference between reactant and product. The QM9x dataset was not designed with any type of molecular dynamics or reaction kinetics in mind, and comparing it to ANI1x and Transition1x for reaction path search is perhaps inappropriate. However, given the ubiquity of QM9 in the literature, it is an important point to convey, that new datasets are required for solving higher order problems in computational chemistry. The Transition1x and ANI1x models drop in performance above 5 eV. Data becomes scarcer at higher energies and consequently, models are less accurate in high energy regions. \gls{dftb} and the ANI1x models have systematic errors in their predictions. The ANI1x models are biased towards high energies in the transition regions as they have not seen the low energy valleys connecting equilibria. The \gls{dftb} potential systematically predicts energies too low. In Table \ref{tab:corrected} the systematic errors are corrected based on the training data. This leads to a lower test error for the ANI1x and \gls{dftb}, but a higher test error for Transition1x underlining that Transition1x models are already very accurate.  

\section{Discussion}
To train models that can properly step in as surrogate potentials for \gls{dft} when running \gls{neb}, it is necessary to have datasets with appropriate data in and around transition-state regions. Finding reaction barriers with \gls{ml} models and \gls{neb} is a non-trivial test. \gls{ml} models, and especially \glspl{nn}, are known to perform poorly for out of distribution tasks \cite{ood1, ood2}. Table \ref{tab:eval_models} illustrates this clearly with results for training and testing \gls{ml} models on various datasets.

Finding reaction barriers with \gls{neb} is a much more demanding test of the models' capabilities. When running \gls{neb}, the \gls{pes} is swept by the path connecting endpoints, and data encountered in the process can diverge wildly from any data seen during testing and training. The model can get caught in even a small region of high error, or it can be thrown off the correct \gls{mep} and make it unable to converge altogether, so the model must be accurate across the entire \gls{pes}. 

The reaction paths are represented by ten images in all reactions. A core strength of \glspl{nn} is their ability to utilize GPUs to evaluate multiple data points at once, and in principle, \gls{neb} can be run with hundreds of images instead of tens at little to no additional cost when using \glspl{nn} as potentials. We ran experiments with high density paths with the rest of the setup fixed but saw no improvement in neither accuracy nor convergence speed. The preconditioning scheme of the \gls{neb} optimizer relies on a sparsely populated path. But this approach could possibly produce robust results by applying other optimizers. 
 
A clear application of this work is as a screening procedure for complex reaction networks. Cheap methods, such as permuting bond order matrices, can be used to automatically generate nodes for entire reaction networks. The individual reactions can be screened fast using the method before recalculating entire reaction networks with expensive methods. Usually this is done with \gls{dftb} \cite{dftb} but running \gls{neb} with \glspl{nn} is faster and more accurate. 

\section{Conclusion}
We have trained state-of-the-art \gls{gnn} potentials on various datasets and used them as surrogate potentials for \gls{dft} when running \gls{neb} for transition-state search. A \gls{mae} of 0.13 eV and \gls{rmse} of 0.24 eV is achieved with the best model, compared against running the same set up with \gls{dft}. The models converge 80.5\% of the time on unseen reactions. We show that expressive models alone are, by no means, sufficient for solving complex tasks in quantum chemistry moving forward, but just as much care has to be put into designing and generating datasets. We tested 3 different datasets: ANI1x, QM9x and Transition1x and only models trained on the latter could reliably solve the transition search task. 

%Development of new datasets for benchmarking and training \gls{ml} models might not be as attractive as research on novel \gls{nn} architectures, but 
Our results show that the future development of the field of \gls{ml} for quantum chemistry stands on two legs -- the completeness of the available data, and the expressiveness of the available models. %We have shown clearly, by training equivalent models on different datasets, that cross-validation on testing data is an inconsequential suggestion for performance on higher order tasks.
%
%The overall performance of even the best \gls{ml} models is no better than the quality of training data. 
Transition1x is a relatively simple dataset that deals with four types of atoms. To apply the results of this paper to general chemistry, larger datasets with more atom types should be produced. Our results indicate that the machine learning approach scales: With the right amount of the right data, accuracies at a sufficient level can be achieved.

\section*{Acknowledgements}
The authors acknowledge support from the Novo Nordisk Foundation (SURE, NNF19OC0057822) and the
European Union’s Horizon 2020 research and innovation program under Grant Agreement No. 957189
(BIG-MAP) and No. 957213 (BATTERY2030PLUS).
\\

\noindent Ole Winther also receives support from Novo Nordisk Foundation through the Center for Basic Machine Learning Research in Life Science (NNF20OC0062606) and the Pioneer Centre for AI, DNRF grant number P1.

\section{Code Availability}
Code for training PaiNN models and running \gls{neb} is available through the repository \href{https://gitlab.com/matschreiner/neuralneb}{https://gitlab.com/matschreiner/neuralneb}.
\balance
\onecolumn

\noindent
\bibliographystyle{unsrtnat}
\bibliography{refs}

\appendix
\newgeometry{top=20pt,bottom=0pt,right=75pt,left=75pt}
\onecolumn

\section{PaiNN Performance On Test Data}
Table displaying results of the models when training and testing on various datasets. In all test set-ups the models that perform best, are models that have been trained on training data from the corresponding dataset. 

\begin{table}[H]
\centering
\begin{tabular}{|ll|ll|ll|}
\hline
              &           & \multicolumn{2}{c|}{Energy {[}eV{]}} & \multicolumn{2}{c|}{Forces {[}eV/Å{]}} \\
              
Trained on    & Tested on & RMSE              & MAE              & RMSE               & MAE               \\
\hline
ANI1x         &  & \textbf{0.04(1)}   &\textbf{0.02(0)}  & \textbf{0.04(0)}   & \textbf{0.01(0)} \\
Transition1x           & ANI1x & 0.35(2)  & 0.22(1) & 0.34(2)  & 0.08(0) \\
QM9x           &  & 3.03(2)  & 2.32(1) & 1.3(7)   & 0.56(2) \\ 
\hline

ANI1x         &    & 0.61(7)  & 0.28(2) & 0.5(1) &0.10(5)  \\
Transition1x           &  Transition1x  & \textbf{0.10(1)}   & \textbf{0.05(0)}  & \textbf{0.10(0)}   & \textbf{0.04(1)} \\
QMx          &    & 2.61(2)  & 1.42(1) & 0.43(0)  & 0.19(0)\\ 
\hline

ANI1x         &    & 0.13(0)   & 0.12(0)  & 0.05(0)   & 0.02(0) \\
Transition1x           &  QM9x  & 0.12(0)   & 0.07(1)  & 0.07(0)   & 0.04(0) \\
QM9x           &    & \textbf{0.04(2)}    & \textbf{0.02(1) } &\textbf{0.01(0)}   &\textbf{0.01(0)} \\

\hline
\end{tabular}
\caption{
Test results of PaiNN models trained on ANI1x, QM9x, Transition1x. We report  \gls{rmse} and \gls{mae} on energy and forces. Force error is the Euclidian distance between the predicted and true force vector.}
\label{tab:eval_models}
\end{table}

\section{Additional Figures}
This section contains the unbounded version of Fig. \ref{fig:histogram} as well as additional plots of \glspl{mep} and \glspl{pes} comparing PaiNN trained on Transition1x with \gls{dft}. 

\begin{figure*}[h]
    \centering
    \includegraphics[scale=0.8]{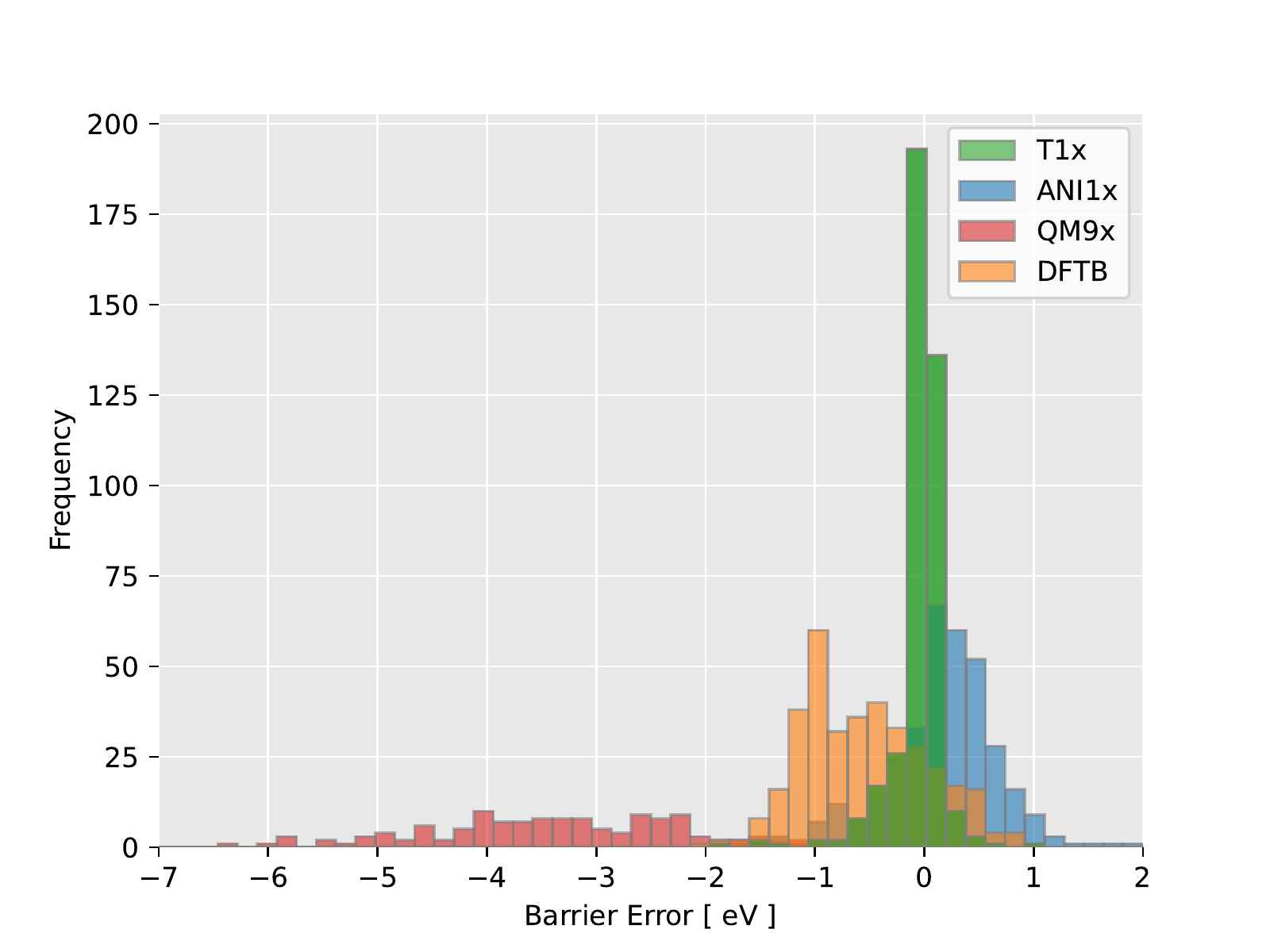}
    \caption{Histogram of barrier errors. The x-axis shows errors between reaction barriers calculated using
Density Functional Theory (DFT) and surrogate potentials for Nudged Elastic Band (NEB). The y-axis shows the frequency of each bin. Green, red and blue display results from PaiNN models trained on Transition1x, QM9x and ANI1x, respectively.
Yellow displays results from Density Functional based Tight Binding (DFTB). The QM9x model has such a low
convergence frequency, and general barrier error, that the model does not show in the plot.}
\end{figure*}

\begin{figure}
  \makebox[\textwidth]{%
    \begin{minipage}[t]{\textwidth}
      \centering
      \includegraphics[width=\textwidth]{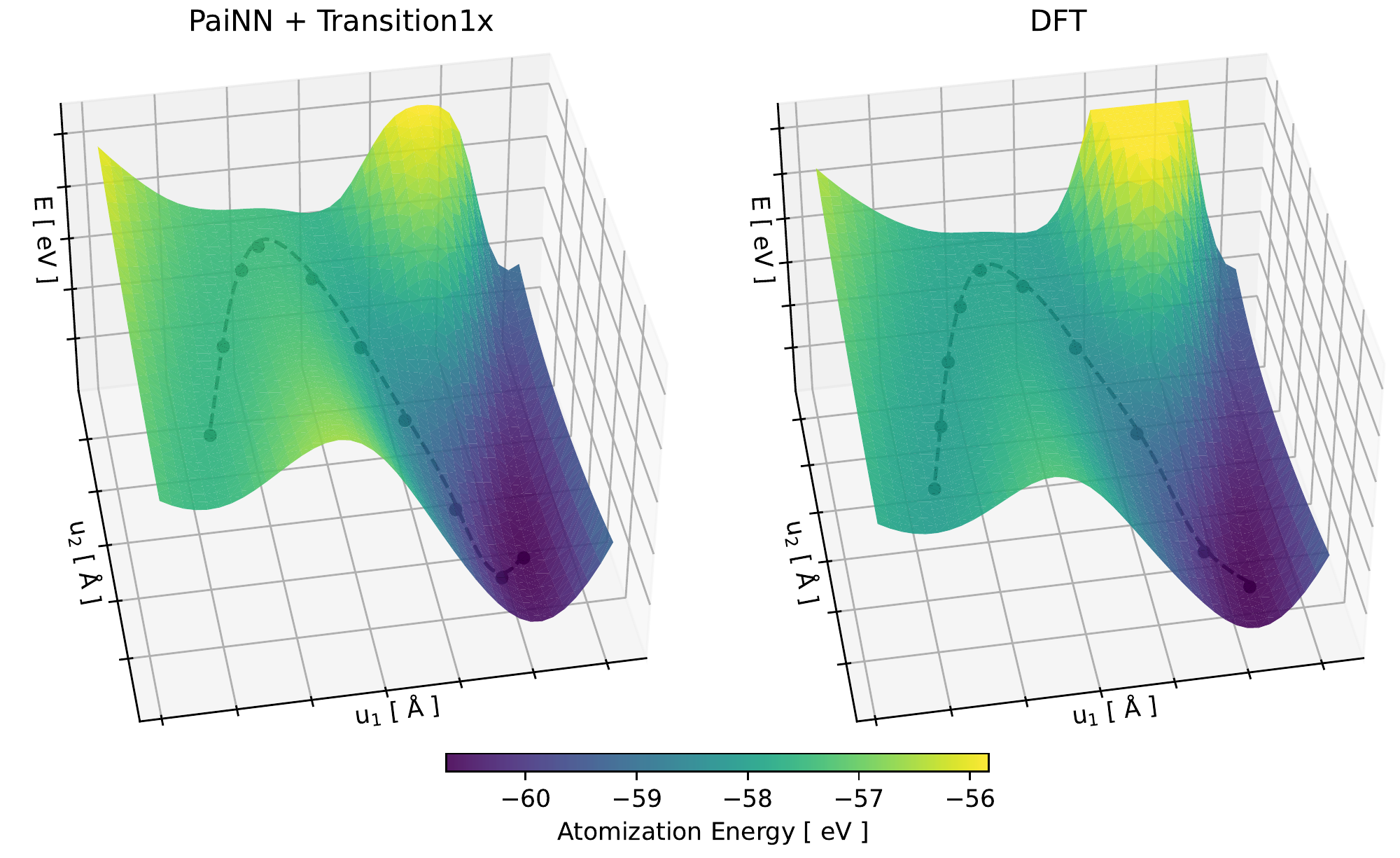}
    \end{minipage}%
  }%
  \caption{Reaction involving C5OH8. The reaction can be seen as a GIF by following  \href{https://figshare.com/articles/figure/Reaction_1210/20382705}{this link}.}
\end{figure}

\begin{figure}
  \makebox[\textwidth]{%
    \begin{minipage}[t]{\textwidth}
      \centering
      \includegraphics[width=\textwidth]{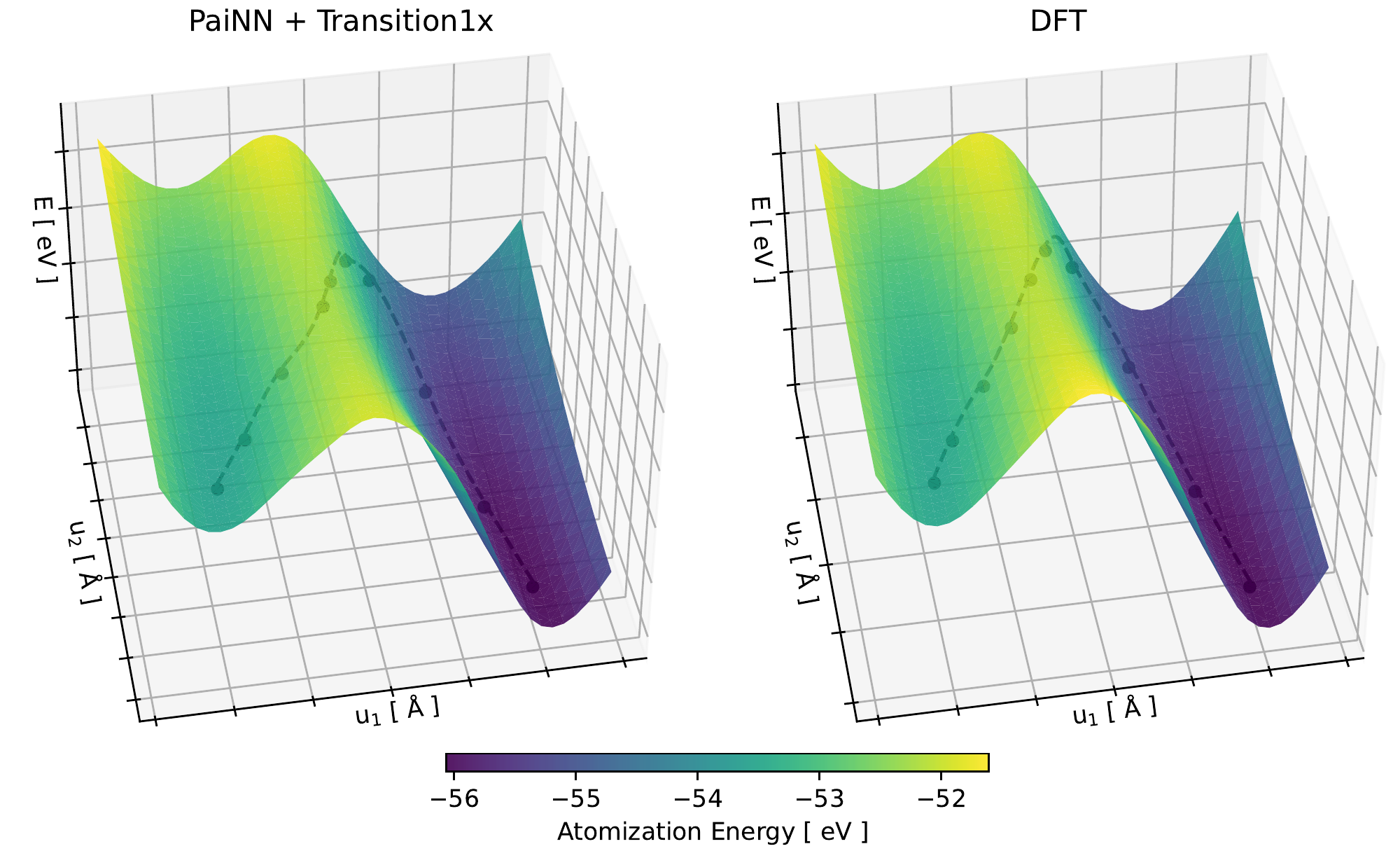}
        \caption{Reaction involving C3NCOH7. The reaction can be seen as a GIF by following  \href{https://figshare.com/articles/figure/Reaction_2456/20382714}{this link}.}
    \end{minipage}%
  }%
  
\end{figure}

\begin{figure}
  \makebox[\textwidth]{%
    \begin{minipage}[t]{\textwidth}
      \centering
      \includegraphics[width=\textwidth]{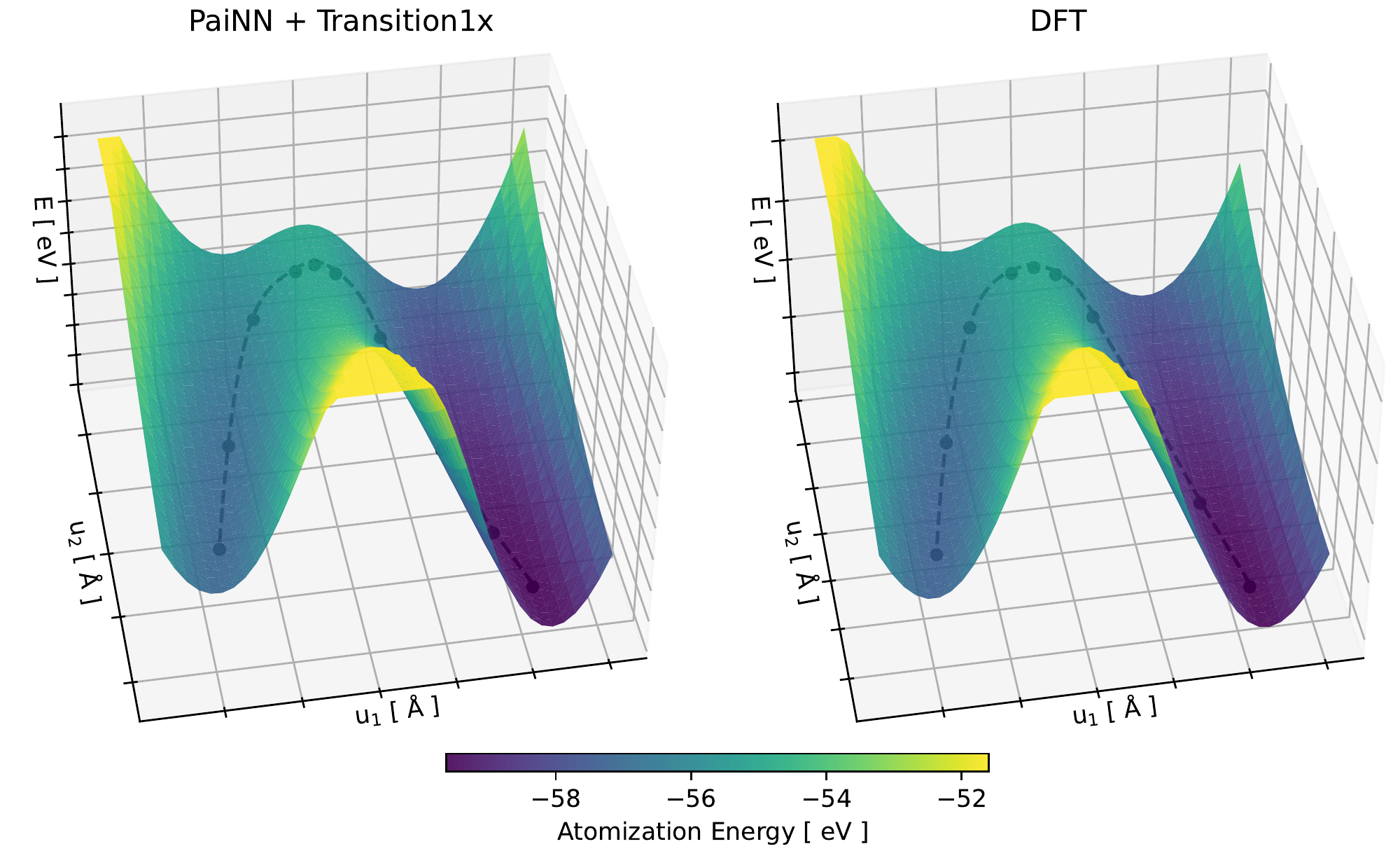}
      \caption{Reaction involving C3NCNH8. The reaction can be seen as a GIF by following  \href{https://figshare.com/articles/figure/Reaction_4617/20382720}{this link}.}
    \end{minipage}%
  }%
\end{figure}

\begin{figure}
  \makebox[\textwidth]{%
    \begin{minipage}[t]{\textwidth}
      \centering
      \includegraphics[width=\textwidth]{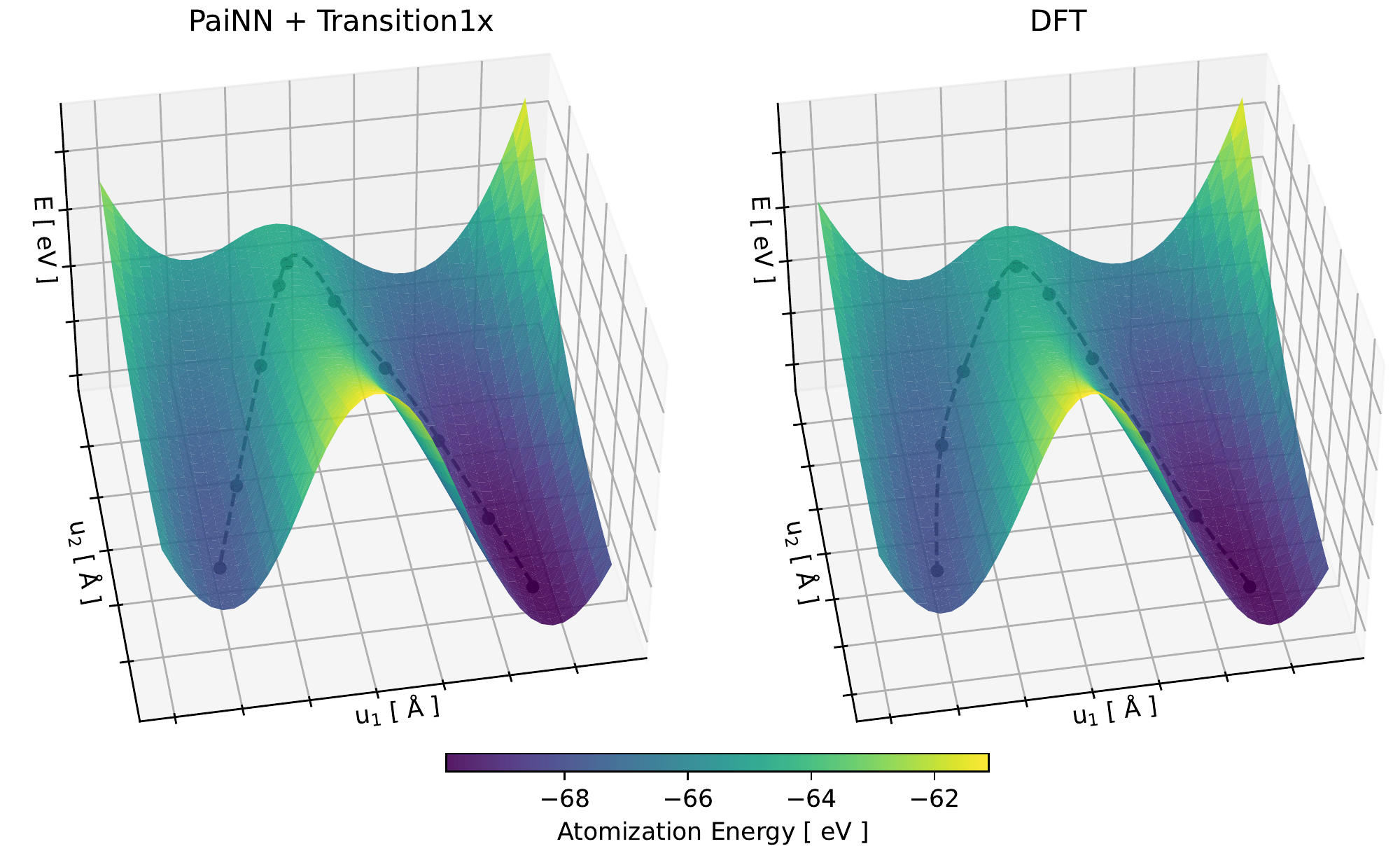}
            \caption{Reaction involving C3NC2OH9. The reaction can be seen as a GIF by following  \href{https://figshare.com/articles/figure/Reaction_6742/20382732}{this link}.}
    \end{minipage}%
  }%
\end{figure}

\end{document}